\documentclass[journal]{IEEEtran}
\usepackage{subfigure}
\usepackage{cite}
\usepackage{hyperref}
\usepackage{csquotes}
\usepackage{balance}
\usepackage{color}
\usepackage{subfigure}
\usepackage{graphicx}
\usepackage{amsmath}
\usepackage{lipsum}

\usepackage{subfigure}
\usepackage{cite}
\usepackage{hyperref}
\usepackage{csquotes}
\usepackage{balance}
\usepackage{color}
\usepackage{subfigure}
\usepackage{graphicx}
\usepackage{tabularx}
\usepackage{lipsum}

\begin{document}

\title{IIFNet: A Fusion based Intelligent Service for Noisy Preamble Detection in 6G}
\author{ Sunder Ali Khowaja$^\dagger$, \IEEEmembership{Senior Member, IEEE}, Kapal Dev$^\dagger$, \IEEEmembership{Senior Member, IEEE}, Parus Khuwaja, Quoc-Viet~Pham, \IEEEmembership{Member, IEEE}, Nawab~Muhammad~Faseeh~Qureshi,~\IEEEmembership{Senior Member,~IEEE}, Paolo Bellavista, \IEEEmembership{Senior Member IEEE}, and Maurizio Magarini
\thanks{$^\dagger$Joint first authors, with equal contributions to this paper}
\thanks{Corresponding authors: Nawab Muhammad Faseeh Qureshi and Kapal Dev}
\thanks{Sunder Ali Khowaja with Department of Telecommunication Engineering, Faculty of Engineering and Technology, University of Sindh, Pakistan. Email: sandar.ali@usindh.edu.pk}
\thanks{Kapal Dev is associated with Department of Computer Science, Munster Technological University, Ireland and the Department of institute of intelligent systems, University of Johannesburg, South Africa, e-mail: (kapal.dev@ieee.org).}
\thanks{Quoc-Viet Pham is with the Korean Southeast Center for the 4th Industrial Revolution Leader Education, Pusan National University, Busan 46241, Republic of Korea (e-mail: vietpq@pusan.ac.kr).}
\thanks{Parus Khuwaja with Institute of Business Administration, University of Sindh, Jamshoro. Email: Parus.khuwaja@usindh.edu.pk}%
\thanks{Nawab Muhammad Faseeh Qureshi is with Department of Computer Education, Sungkyunkwan University, Seoul, Korea. (E-mail:faseeh@skku.edu).}
\thanks{Paolo Bellavista is associated with University of Bologna, Italy. e-mail: (paolo.bellavista@unibo.it).}
\thanks{Maurizio Magarini is associated with Dipartimento di Elettronica, Informazione e Bioingegneria, Politecnico di Milano,Italy. e-mail: (maurizio.magarini@polimi.it).}
}

%


\maketitle

\begin{abstract}
 	
In this article, we present our vision of preamble detection in a physical random access channel for next-generation (Next-G) networks using machine learning techniques. Preamble detection is performed to maintain communication and synchronization between devices of the Internet of Everything (IoE) and next-generation nodes. Considering the scalability and traffic density, Next-G networks have to deal with preambles corrupted by noise due to channel characteristics or environmental constraints. We show that when injecting \emph{15\%} random noise, the detection performance degrades to \emph{48\%}. We propose an informative instance-based fusion network (IIFNet) to cope with random noise and to improve detection performance, simultaneously. A novel sampling strategy for selecting informative instances from feature spaces has also been explored to improve detection performance. The proposed IIFNet is tested on a real dataset for preamble detection that was collected with the help of a reputable commercial company. 
\end{abstract} 

\section{Introduction}\label{sec:intro}
The 5G communication system stacked the Internet of Things (IoT) on top of the 4G mobile broadband to attain enhanced mobile broadband (eMBB), accordingly. The eMBB due to IoT implicitly ensures the inclusion of ultra-reliable and low-latency communication (uRLLC), and massive machine type communication (mMTC), respectively. Although the worldwide deployment of 5G has begun in 2020, it is assumed that the said communication system will not be able to meet the requirements for Next-G communication systems \cite{Du2020veh}. In this regard, researchers have begun working on 6G standards to overcome some of the associated limitations. \par
A report from Gartner\footnote{http://www.gartner.com/newsroom/id/2684616} states that by the end of 2025 the connection of more than 30 billion devices to the communication network will be observed. The surge in connection devices requires low-latency communication to be more reliable and accommodation of mMTC to be more scalable in comparison to 5G.  The vision of Next-G cellular communication via 6G is to expand the coverage and boundaries of the aforementioned services, with Internet of Everything (IoE) and artificial intelligence (AI) as its key enabling technologies \cite{Du2020veh}. 
Similarly, user-experienced data rate, traffic capacity, spectrum efficiency, energy efficiency, connection density, mobility support, coverage, security capacity, and cost efficiency are assumed to improve with the emergence of Next-G networks \cite{Du2020veh}. An example of different applications in relation to Next-G networks is shown in Fig.~\ref{Fig1}. The Next-G allows the users to have ubiquitous services, consistent and generalized mobility irrespective of the underlying communication medium or transport technology\footnote{https://www.igi-global.com/dictionary/modelling-quality-and-pricing-in-next-generation-telecom-networks/20320}. The idea is illustrated in the aforementioned figure, such that the services are provided and users are accommodated irrespective of the coverage area or fixed infrastructure. The connectivity needs to be ensured through devices that can connect to the Internet or are capable of establishing a connection with communication satellites. The applications to Next-G communication systems include but are not limited to smart grids, Industry 5.0, smart homes, smart transportation systems, disaster recovery systems, and remote tourism \cite{Dev2021}. \\
As Next-G networks need to support more mMTC devices to fulfil the scalability needs, random access (RA) procedure is of vital importance as it ensures the uplink synchronization from IoE related devices to the base station. We will refer to the 5G standard for further explanation, as the standards for Next-G have not been proposed yet. 
The transmission of random access preambles from IoE devices to generation node B (gNB) is carried out through the physical random access channel (PRACH) \cite{wu2020efficient}. The gNB detects the preambles while processing the PRACH signals to assign the timing advance and the preamble ID (PID). The signal information is then transmitted in a response message to IoE device for adjusting the transmission time and establishing synchronization with gNB\footnote{3GPP Release 15  and Release 16}. In case of PRACH process failure, a preamble needs to be sent again after a pre-defined amount of time that degrades the performance and suffers from unnecessary delay \cite{Sharma2020}.

\begin{figure*}[ht!]
\centering
  \includegraphics[width=\linewidth]{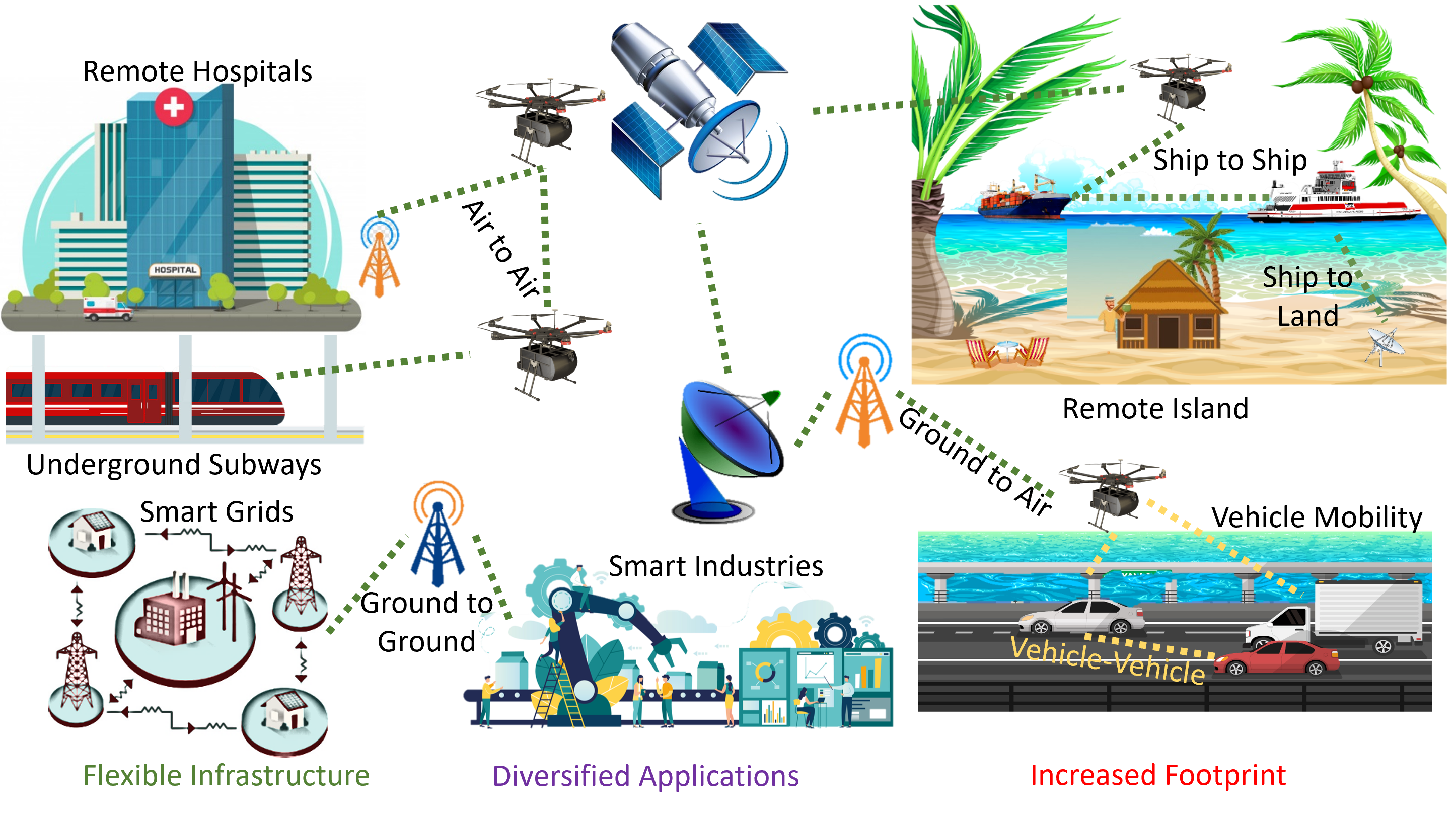}
  \caption{An overview of Next-G communication networks.}
  \label{Fig1}
\end{figure*}

The existing works for preamble detection have shown promising results \cite{yang2020mixed}\cite{zhang2020tara} by achieving a detection rate of 99$\%$ using threshold-based techniques on signal-to-noise ratio. However, these techniques fail to generalize the performance on large number of devices due to the false peaks \cite{Sharma2020, Modina2019}. Researchers have since then moved to machine learning approaches for preamble detection \cite{Mostafa2021}. 
The problem with the existing methods employing machine learning is that they do not consider the random noise problem that can affect the data collection process itself. This random noise can manipulate the training process to classify false peaks as true ones. Detection of false peaks not only increases latency, but also affects network efficiency and scalability. To the best of our knowledge, the random noise problem has not been dealt in PRACH processing.\par

In this study, we collected primary data from a well-known company working in the field of communication systems for preamble detection in accordance with the new 5G radio systems and 3GPP technical specification group radio access network NR physical channels and modulation\footnote{3rd GPP Technical Specification Group Radio Access Network NR Physical Channels and Modulation, 38.211, 2019}. Data are gathered specifically when an IoT device transmits preambles to request an uplink allocation to the gNB. The data are then injected with random additive white Gaussian noise levels varying between 5 - 15$\%$. We applied several shallow- and deep-learning approaches to show the effect of random noise on false-peak detection, respectively. In this regard, we propose an informative instance-based fusion network (IIFNet) for not only detecting the preambles accurately but also to deal with the random noise problem in PRACH processing. The contributions of this work are stated below:\par

\begin{enumerate}
	\item Collection of primary data for preamble detection in compliance to 5G-new radio systems.

	\item A new sampling strategy is proposed to select the most informative instances.

	\item A novel fusion network, IIFNet is proposed for reliable detection of preambles in noisy environment.
	
	\item State-of-the-art results for preamble detection on noisy data have been reported.
\end{enumerate}

\vspace{\baselineskip}


\begin{figure*}[h]
\centering
  \includegraphics[width=\linewidth]{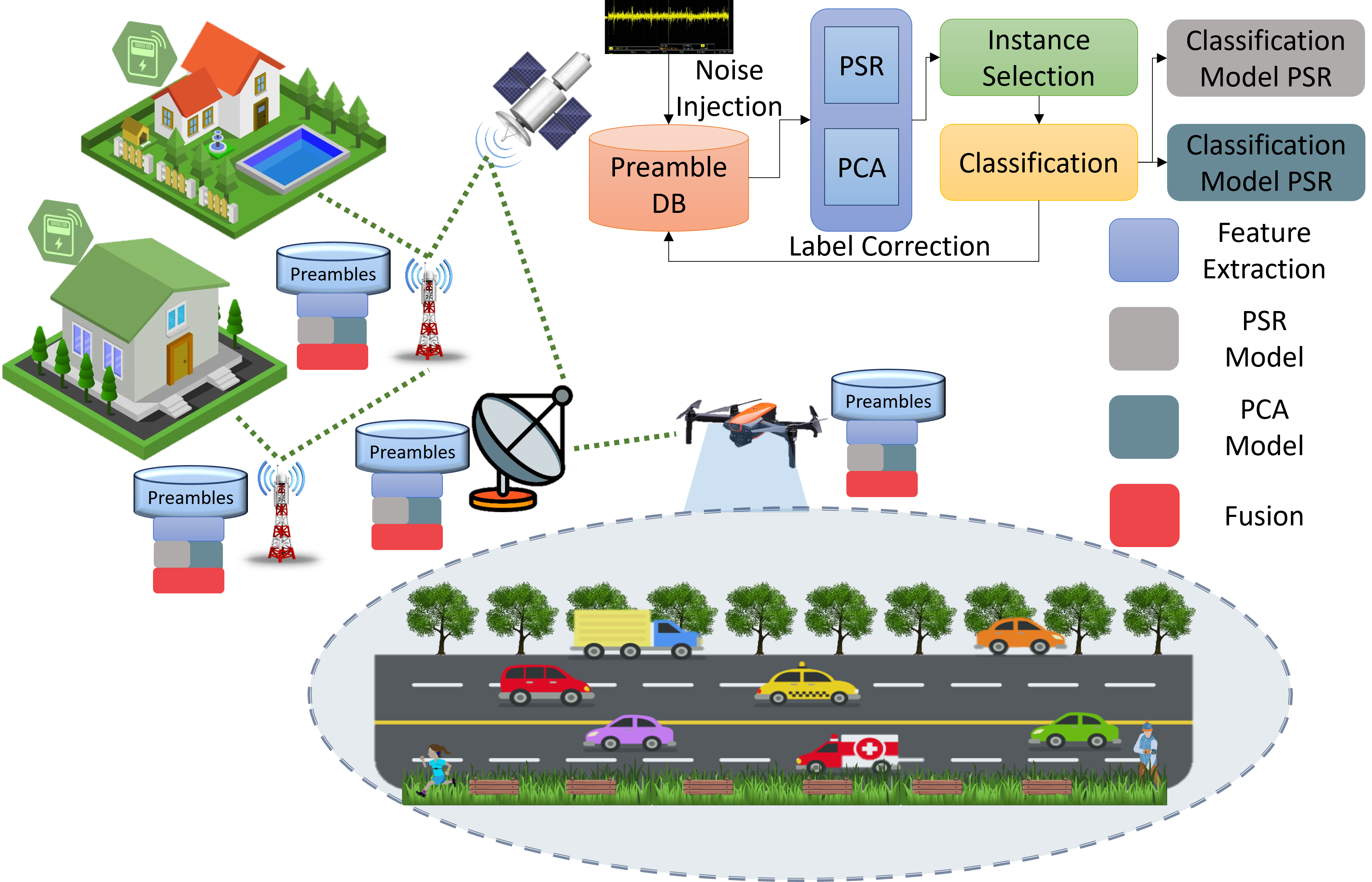}
  \caption{Proposed IIFNet framework for preamble detection in 6G and beyond communication networks.}
  \label{Fig2}
\end{figure*}


\section{State-of-the-Art Studies}\label{sec:BG}
Telecommunication industry is rapidly growing and
integrating with wide array of technologies due to the emergence of 6G communication systems. The 6G services are supported by enhanced eMBB, lower uRLLC, and more mMTC. For consolidating a brief review, we consider 5G-new radio (NR) to be a reference point. The authors in \cite{dahlman20185g} discussed the importance of optimizing downlink resource slicing and its impact on both the uRLLC and the eMBB with respect to 5G-NR systems. Amongst many characteristics of downlink resources, the study emphasized the importance of preamble detection/design for improving the network performance. Thanks to several favorable characteristics like constant amplitude and low cross-correlation, Zadoff-Chu (ZC) sequences are employed to generate random preambles in LTE and 5G-NR \cite{yang2020mixed}\cite{lin2016random}. As one of the seminal works on the design of 5G-NR for narrowband IoT (NB-IoT), a single-tone waveform and frequency hopping NB-IoT PRACH (NPRACH) was proposed in \cite{wu2020efficient}. The aforementioned design had the merit of being nearly zero peak-to-average power ratio, thus suitable for NB-IoT systems with strict requirements of long battery lifetime, low cost, and extensive coverage. When multiple IoT devices try to access the network simultaneously, the NR may receive superimposed NPRACH preambles, and further need to detect them that are received from different IoT devices. In this network configuration, the work in \cite{zhang2020tara} solves the preamble detection problem to obtain the optimal Neyman-Pearson threshold in NB-IoT systems. The work also focused on estimating the time-of-arrival and residual frequency offset of the detected IoT devices. The work in \cite{zhen2018random} exploits the difference in time-of-arrivals of collided preambles to further improve the performance of NB-IoT systems.

As AI is one of the key enabling technologies for 6G
communication systems, many researchers have adapted it for improving the preamble detection process. In \cite{Mostafa2021} a deep learning-based method was developed for the decoding of preambles. The study aggregated to ZC-sequences to mimic the effect of massive IoT devices for the use-case of 5G systems and designed two separate decoders. The first was based on threshold-based measurements, while the second used a deep learning approach to detect the preambles accordingly. A deep neural network architecture was developed in \cite{ding2020deep} for preamble collision detection in grant-free random access MIMO systems. The key idea of the work was that only nearby base stations of a collided user are used, instead of all base stations in the network. Simulation results show that their deep model for base station clustering yields a higher achievable rate compared to the baseline alternatives. The study \cite{Modina2019} proposed the use of shallow approaches to detect the preamble in 5G systems. It should be noted that all the deep learning based approaches are applied to the data generated using synthetic means. Moreover, none of the said strategies has considered the random noise problem that significantly degrades the preamble detection performance. 

Many studies have proposed the use of majority voting and consensus voting methods \cite{Samami2020} to remove random noise, but they will not work in the field of communication systems, as the peaks are detected at a much lower rate compared to the false peaks, and the aforementioned methods tend to eliminate the preambles if the sampling frequency is lower. To the best of our knowledge, this is the first study to deal with the random noise problem for preamble detection in the PRACH process.

\section{Proposed Method}

It is well established that noise is an integral part of the communication system that needs to be modeled to remove any ambiguity in the detection process. The aim of this work is to design a framework that is capable of accurate detection even if the samples are corrupted by the random noise at the physical layer. In this regard, we propose an informative instance-based fusion network (IIFNet) to improve the detection process as shown in Fig.~\ref{Fig2}. To the best of our knowledge, such sampling strategy for selecting the most informative instances in the context of preamble detection has not been proposed before. The preambles are collected from IoT devices
and stored in a preamble database. We inject the random
Gaussian noise into the preamble database to create distorted data as should be in real-life scenarios. We then transform the raw feature space, i.e., amplitude, variance, threshold, and SNR to phase space reconstruction (PSR) \cite{Khuwaja2020}, and principal component analysis (PCA).



The intuition behind adopting the said feature transformation techniques is to cover a wide range of feature engineering spectrum. The PSR projects the lower dimensional data onto higher dimensions, whereas the PCA takes the inverse approach. The higher dimensions in PSR would sense the data impurity through information gains such as distance measures, whereas the PCA would only consider the feature variables that could explain the major portion of the variance that filters the data in an intrinsic manner. Moreover, PCA has been used extensively for pre-processing and denoising the data, respectively. We then select a portion of informative instances from the feature transformation technique that defines the entire feature space and train individual classifiers. The trained classifiers will then be used to label the remaining instances, and the same process will be iterated over until all instances are labeled accordingly. Once the classification models are trained, we use them to classify the preambles. As the classification models are trained individually for both feature transformation techniques, we perform decision level fusion (DLF) \cite{Khuwaja2020} on the predicted labels to further improve the detection process. These trained models and fusion strategies are attached to each gNB, evolved node B (eNB), or base stations to distinguish between true and false peaks accordingly.

\subsection{Random Noise Injection}
Artificial noise injection is a common method in the research studies to corrupt the data samples. Random Gaussian noise (RGN) injects the noise by changing the labels of each sample with a varying probability ranging from 0 to 1. Assuming that the power level is quite low in Next-G communication systems, we considered a zero-mean additive white Gaussian noise (AWGN) for injection. As the mean is zero, we use the square root of the mean power level of the dataset to be used as the standard deviation. We create a Gaussian distribution and corrupt a specific portion of the dataset, respectively. During the ablation study, it was observed that a simple threshold-based technique can accurately differentiate between the true and false peaks. However, adding just 5$\%$ noise represents a difficult problem for many sophisticated classification algorithms, since accuracy decreases by more than 7$\%$ at times. Further increase in the noise level leads to larger variations and exponential degradation in detection performance.

\subsection{Phase Space Reconstruction}
The PSR has been extensively used for the studies related to non-linear dynamics. It projects a one-dimensional vector to  \( k \) -dimensional space with respect to the delay  \( T \). Although, the projected vector in  \( k \) -dimensional space is equivalent to one-dimensional vector, it considers each higher dimensional space as a probability distribution. These distributions are considered as a meaningful feature \cite{Khuwaja2020}. In this study, we assume that each data point is an independent and identically distributed (IID) variable; therefore, we set the time lag for PSR as 1. This suggests that for each embedding dimension 1 $ \ldots $   \( k \), the time lag is added by 1, accordingly. The projection of IID data onto higher dimensions in such a way leads to an evolutionary trajectory in embedding space. The PSR embeddings will be used to calculate information distances.

\subsection{Principal Component Analysis}

The PCA, unlike PSR, projects time series data to lower dimensions by choosing coefficients based on their variance. In this study, we use PCA as one of the feature transformation techniques. In comparison to PSR, we use embedding dimensions in PCA as well; however, for PCA, the selected dimensions are less than or equal to the actual ones. The PCA computes a covariance matrix that comprises Eigenvalues  
which are sorted in descending order accordingly. Considering that the raw feature comprises 4 dimensions, we get 4 eigenvalues. We need to select fewer than four eigenvalues to project the original feature space to lower dimensions. 

\subsection{Instance Selection}
Existing studies have used majority or consensus filtering based approaches to deal with the random noise problem. The majority filtering is quite sensitive to the data itself and provides a lot of false positives whereas the consensus filtering is conservative and mislabels the actual instances as noise. We take a different path by leveraging the feature transformation techniques used in this study. We select the informative instances, which, in turn, will let the system correct the label if needed. In the past, the selection of informative instances has been carried out using pool-based uncertainty sampling and query by committees. This study considers the density and uncertainty based samples to select the informative instances. The term “density” is defined as the similarity between other examples and the similar ones within the same cluster, as the clusters are very large and the density values are quite close to one another. The second is the skewed distribution of the clusters that creates a bias to a particular label. To overcome the aforementioned problems, we use nearest neighbor based density measure \cite{Samami2020} for computing the similarity between an input instance and its nearest neighbors to quantify its uncertainty. We assume that the density quantifies the uncertainty of a given instance suggesting that higher density maximizes uncertainty; therefore, our proposed informative instance selection characterizes both the density and the uncertainty, simultaneously. The sampling strategy suggests that a sample with higher uncertainty and density will be given more weightage in comparison to other samples at each learning cycle. The process of sampling strategy and training is defined in the following steps:

\begin{itemize}
    \item First, we partition the data in train and validation sets. We select 10\% data for the training process.
    \item A base classifier is initially trained using the train set.
    \item The trained model is then used to label the instances in the validation set.
    \item The instances with high density and uncertainty are sampled accordingly.
    \item The selected instances are augmented to the training data, the said samples are removed from validation set, and the classifier is re-trained on the training set with augmented samples.
    \item The process will be carried out repeatedly until there are no instances left in the validation set.   
\end{itemize}

	
			
			
			
		

\subsection{Classification and Fusion}
A good classification algorithm can help to achieve better recognition performance. Although the proposed IIFNet can accommodate any classification algorithm, but to evaluate its performance, we compare eight techniques that include decision trees, support vector machines, adaptive boosting, extreme gradient boosting, K-nearest neighbor, extreme learning machines, 1-D CNN, and long short-term memory networks. The six former classifiers belong to the shallow learning category of classification methods. Each of the chosen classifier belongs to a different family, thus, proves the applicability with IIFNet and also covers wide range of spectrums. The latter two belong to the deep learning category, where the paradigm has been shifted recently concerning the classification and detection strategies in the context of next generation communication systems. The importance of deep learning methods is justified from the works reviewed in literature (Section II). The classifiers are trained in conjunction with automatic hyperparameter optimization, suggesting that each classifier is trained/evaluated several times with varying parameters and the one attaining the best performance will be reported in the results.

In the machine learning literature, studies tend to combine the results from multiple classifiers in order to boost the recognition performance. As we are training individual classifiers for the PSR and PCA transformations in the IIFNet, we can naturally combine their predictions to further improve the preamble detection performance. In this study, we employ weighted averaging, and meta-learning using Naïve Bayes to fuse the decision labels, accordingly. The meta-learning strategy refers to a machine learning algorithm stacked onto the decisions of multiple base classification methods. IIFNet combines the two streams, that is, the best classifier results using PSR features and the best classifier results using PCA features. A Naïve Bayes classifier is trained on the probability of occurrences given the samples from two-streams mentioned above to yield the final output. This process is referred to as meta-learning using Naïve Bayes, in this study. \par

\section{Experimental Setup and Results}
In this section, we first provide a brief introduction to our primary data. We then present details regarding the experimental setup and parameters regarding the proposed IIFNet framework. We first provide the results related to our problem formulation, i.e., the results without and with noise injection. The results provide a basis for the IIFNet framework and will also be used as a baseline to evaluate the performance.\par

\subsection{Dataset}
The primary dataset for preamble detection has been collected with the collaboration of reputed commercial company based in Italy that operates in the field of wireless communication and specializes in small cell solutions, LTE / HSPA + and C-RAN, accordingly. The parameter setting for the data collection process is shown in Table \ref{tab:Tab1}. The data set was collected according to the 3GPP standards for the radio access network. It includes the amplitude, threshold, variance, and SNR values with their corresponding labels, i.e., FalsePeak and Peak. The data comprises more than 100,000 samples out of which, 92000 samples and 8000 samples correspond to FalsePeak and Peak, respectively. The dataset has varying scales; therefore, we performed mean normalization before providing the data to the IIFNet framework.

\begin{table}[]
\centering
\caption{Parameter settings of the primary data collected.}
\label{tab:Tab1}
\begin{tabular}{|l|l|}
\hline
Parameter & Values \\ \hline
System Bandwidth & 20 MHz \\ \hline
PRACH Format & 0 \\ \hline
Channel & AWN/ETU 70 \\ \hline
Doppler & 0/200 Hz \\ \hline
Rx Antennas & 2 \\ \hline
Ncs & 13 \\ \hline
\begin{tabular}[c]{@{}l@{}}Data Subscriber\\ Spacing\end{tabular} & 1.25 KHz \\ \hline
\# of Sequences & 1000 \\ \hline
Frame Length & 800 $\mu$sec \\ \hline
SNR & 10 dB \\ \hline
\end{tabular}
\end{table}

\subsection{Setup}
All experiments in this study are carried out using MATLAB R2018b on a PC clocked at 3GHz, and RAM of 32 GB. We injected the data with 5$\%$, 10$\%$, and 15$\%$  noise levels. The PSR method requires the time lag and number of embedding dimensions. It was already mentioned that due to IID characteristics, the time lag was selected to 1. We empirically select the number of embedding dimensions for PSR and PCA to be 7 and 2, respectively. We use the hyperparameter toolbox to dynamically select the parameters for each classifier based on the best performance. The dataset was split into 70$\%$ and 30$\%$  for training and testing, respectively. For fair results, we repeated the experiments five times by randomly selecting the proportion of dataset, the reported results in the subsequent sections are averaged, accordingly. As the primary dataset is highly imbalanced, we report performance in terms of accuracy and the F1 score. 

\subsection{Experimental Results}
We present the results without and with varying noise levels using different classifiers in Table 2. It was noticed that without noise, each of the classifiers yield 100$\%$  accuracy for preamble detection. However, when adding noise with varying levels, the accuracy decreases to $ \sim $ 66$\%$  which supports our problem formulation. Amongst all the classifiers, extreme learning machine (ELM) yields the best performance. Therefore, we will use ELM for analysis regarding the selection for value of  \( J \) representing the number of highly dense and uncertain samples. The sensitivity analysis for the selection of  \( J \)  is shown in Fig.~\ref{Fig5}. The results are carried out using selection of informative instances on PSR features and learning through ELM classifier while injecting 15$\%$  noise. The best results were achieved with  \( J=20 \). The results show an improvement in the F1 score by 7.86$\%$ while learning only from 150 instances. This proves our assumption that the proposed sampling strategy makes the classifier noise resilient. We report the F1-scores for each classifier trained on PSR and PCA features using IIFNet in Table \ref{tab:my-table}. The findings show that the PSR performs better than the PCA features, in general. The best performing classifier in terms of preamble detection is ELM for both features; therefore, we use the ELM predictions from both PSR and PCA to perform the DLF. It is also observed that the PSR and PCA features complement each other when used in a fusion scheme, as IIFNet can improve the preamble detection performance by approximately 33$\%$, which is quite remarkable. It should also be noted that the proposed framework (without fusion) improves the preamble detection performance for each classifier, in general.






\begin{figure}[h]
\centering
  \includegraphics[width=0.45\textwidth]{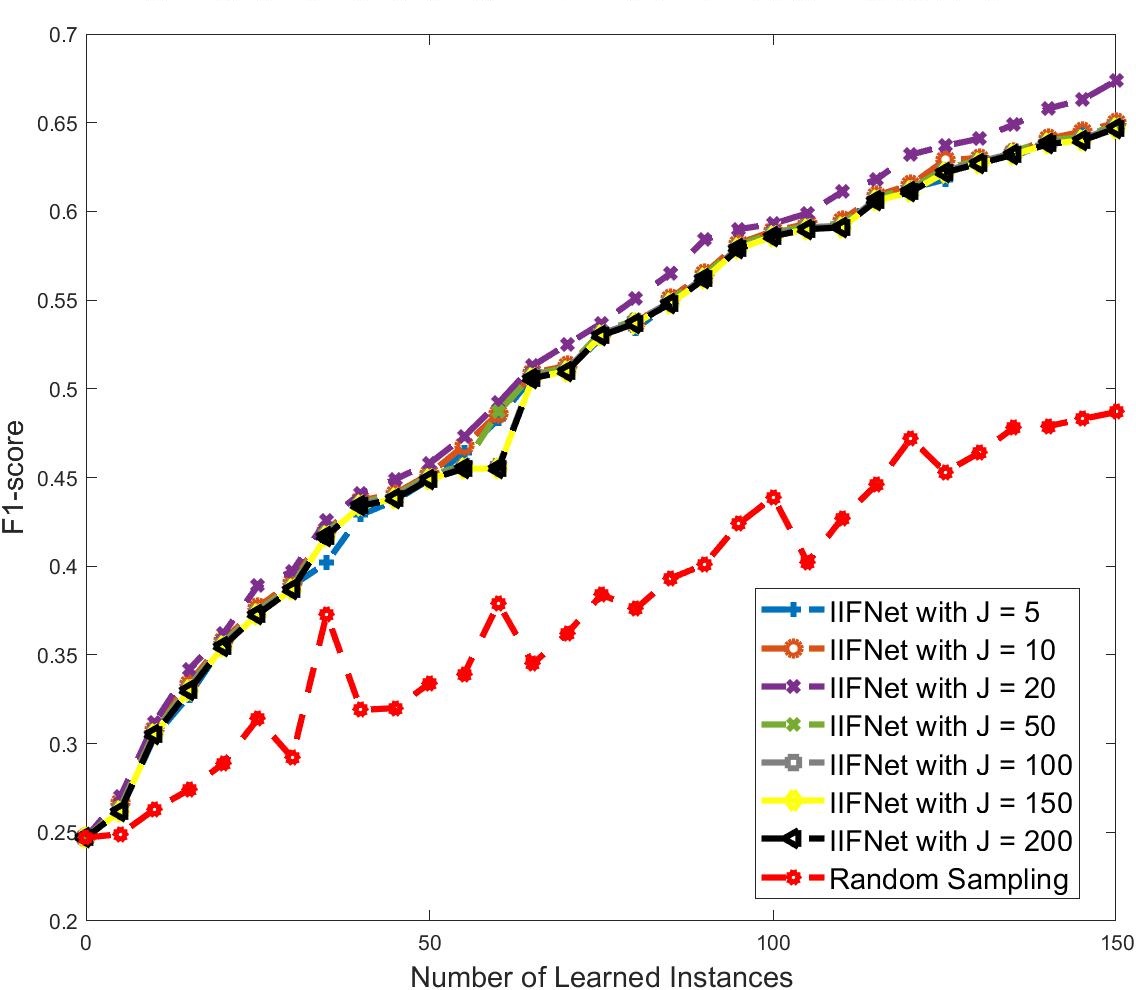}
  \caption{Results of the proposed sampling strategy in IIFNet on the collected dataset, with $J$ varying from 5 to 200. The results are obtained using PSR features trained with ELM.}
  \label{Fig5}
\end{figure}


\begin{table}[]
\centering
\caption{F1-Scores for Each Classifier using Sampling Strategy with PSR, PCA, and Proposed IIFNet}
\label{tab:my-table}
\begin{tabular}{|lccc|}
\hline
\multicolumn{1}{|l|}{Classification Method} & \multicolumn{1}{l|}{5\% Noise} & \multicolumn{1}{l|}{10\% Noise} & \multicolumn{1}{l|}{15\% Noise} \\ \hline
\multicolumn{4}{|c|}{PSR Features}                                                                                         \\ \hline
\multicolumn{1}{|l|}{Decision Trees}                  & \multicolumn{1}{c|}{0.9337} & \multicolumn{1}{c|}{0.8943} & 0.7236 \\ \hline
\multicolumn{1}{|l|}{Support Vector Machines}         & \multicolumn{1}{c|}{0.9786} & \multicolumn{1}{c|}{0.9274} & 0.8284 \\ \hline
\multicolumn{1}{|l|}{Adaptive Boosting}               & \multicolumn{1}{c|}{0.9449} & \multicolumn{1}{c|}{0.8524} & 0.7623 \\ \hline
\multicolumn{1}{|l|}{Extreme Gradient Boosting}       & \multicolumn{1}{c|}{0.9562} & \multicolumn{1}{c|}{0.8978} & 0.7845 \\ \hline
\multicolumn{1}{|l|}{K-Nearest Neighbor}              & \multicolumn{1}{c|}{0.9300} & \multicolumn{1}{c|}{0.8876} & 0.7669 \\ \hline
\multicolumn{1}{|l|}{Extreme Learning Machines}       & \multicolumn{1}{c|}{0.9926} & \multicolumn{1}{c|}{0.9417} & 0.8592 \\ \hline
\multicolumn{1}{|l|}{1-D Convolution Neural Networks} & \multicolumn{1}{c|}{0.9467} & \multicolumn{1}{c|}{0.8725} & 0.7934 \\ \hline
\multicolumn{1}{|l|}{Long-Short Term Memory Networks} & \multicolumn{1}{c|}{0.9624} & \multicolumn{1}{c|}{0.9148} & 0.8377 \\ \hline
\multicolumn{4}{|c|}{PCA Features}                                                                                         \\ \hline
\multicolumn{1}{|l|}{Decision Trees}                  & \multicolumn{1}{c|}{0.9293} & \multicolumn{1}{c|}{0.8465} & 0.6922 \\ \hline
\multicolumn{1}{|l|}{Support Vector Machines}         & \multicolumn{1}{c|}{0.9367} & \multicolumn{1}{c|}{0.8923} & 0.7847 \\ \hline
\multicolumn{1}{|l|}{Adaptive Boosting}               & \multicolumn{1}{c|}{0.9148} & \multicolumn{1}{c|}{0.8229} & 0.7116 \\ \hline
\multicolumn{1}{|l|}{Extreme Gradient Boosting}       & \multicolumn{1}{c|}{0.9209} & \multicolumn{1}{c|}{0.8577} & 0.7342 \\ \hline
\multicolumn{1}{|l|}{K-Nearest Neighbor}              & \multicolumn{1}{c|}{0.9134} & \multicolumn{1}{c|}{0.8241} & 0.7192 \\ \hline
\multicolumn{1}{|l|}{Extreme Learning Machines}       & \multicolumn{1}{c|}{0.9408} & \multicolumn{1}{c|}{0.9066} & 0.8138 \\ \hline
\multicolumn{1}{|l|}{1-D Convolution Neural Networks} & \multicolumn{1}{c|}{0.9223} & \multicolumn{1}{c|}{0.8591} & 0.7370 \\ \hline
\multicolumn{1}{|l|}{Long-Short Term Memory Networks} & \multicolumn{1}{c|}{0.9321} & \multicolumn{1}{c|}{0.8937} & 0.7875 \\ \hline
\multicolumn{4}{|c|}{IIFNet}                                                                                               \\ \hline
\multicolumn{1}{|l|}{Weighted Averaging}              & \multicolumn{1}{c|}{1.00}   & \multicolumn{1}{c|}{0.9723} & 0.8917 \\ \hline
\multicolumn{1}{|l|}{Meta-learning (NB)}              & \multicolumn{1}{c|}{1.00}   & \multicolumn{1}{c|}{0.9837} & 0.9274 \\ \hline
\end{tabular}
\end{table}

\section{Open Issues, Challenges, and Future Directions}
\begin{itemize}
\item \textbf{\textit{Large-scale Fading}}: Similarly to noise, fading also plays an important role in the degradation of a signal. Particularly, in Next-G networks, cell-free massive MIMO are used. Fig.~\ref{Fig1} illustrates that multiple antennas are co-located in a cell-free massive MIMO; therefore, the signal undergoes large-scale fading at each antenna in a different manner. In such cases, the preamble detection system might cause delays, i.e., re-sending preambles after a specific amount of time. We believe that IIFNet can be used for the problem of fading in an effective manner if some data can be collected for such a phenomenon. It is currently one of the future directives of this study.
\item \textbf{\textit{Macro Diversity and Spatial Sparsity}}: Next-G networks propose the use of non-terrestrial networks (NTNs) in combination with infrastructure-based systems. Macro diversity and spatial sparsity are missing from infrastructure-based systems, as devices not close to the antenna or lack of strong signals are ignored. As illustrated in Fig.~\ref{Fig1}, the NTN overcomes this limitation but adds the challenge of asynchronous reception, which leads to the reception of signals at distributed antennas, thus introducing propagation delays for preamble detection. IIFNet can be combined with compensation schemes or timing estimation methods to reduce the effect of asynchronous reception on preamble detection performance.
\item \textbf{\textit{Devise Diversity}}: Fig.~\ref{Fig1} illustrates the use of satellites, high altitude platform stations, and UAVs as NTN nodes. These are often referred to as 3D networks. The preamble detection method can be affected with the diversity of device, type, antenna design, line-of-sight availability at high altitudes, and characteristic disparities. Acquiring data separately for each of the NTN devices may not be possible at first. In this regard, domain adaptation techniques could be exploited in conjunction with IIFNet to improve the detection performance in Next-G 3D Networks.
\item \textbf{\textit{Energy Efficiency}}: At times when communication infrastructure is not available, UAVs can be deployed to collect data for random access in the capacity of MIMO gateways, as shown in Fig.~\ref{Fig1}. Since UAVs have resource constraints, such as energy efficiency, battery capacity, and computation resources, it is crucial that the detection techniques employed are lightweight. We have shown that IIFNet is not confined to a specific classification method; therefore, lightweight detection techniques can be used to further reduce the computational complexity and increase the energy efficiency, accordingly.
\item \textbf{\textit{Anomaly Detection}}: With the diversified NTN nodes and devise diversity, the system might detect the probable unwanted broadcasts as a result from some anomalous equipment behavior. These broadcast can occur while initiating a request to gNB/eNB. As IIFNet uses PSRs and PCAs to transform the feature space, it can naturally be extended to detect the device type as well, provided that the data embed the device-type information. On the basis of the device identification, anomalous behavior can also be detected, concurrently.
\end{itemize}

\section{Conclusion}
This paper focuses on the detection of the preamble for the PRACH process in Next-G communication networks. We showed that by injecting noise into the data, the performance degrades to 48.75$\%$  in terms of the F1 score. In this regard, we proposed IIFNet that uses a novel sampling strategy to select the informative instances from the transformed feature spaces. Furthermore, the use of two different feature transformation methods allows us to perform DLF on classifier predictions so that the detection performance could be further improved. IIFNet has been reported to improve the F1 score by 33.2$\%$ from baseline using the ELM classifier and Naïve Bayes as a meta-learner for DLF. We assume that reducing the false detection of peaks helps the communication systems such as 6G to maintain the high throughput as it reduces the delay in terms of requests.

\section*{Acknowledgment}

The authors thank all contributors who shared data with us for this analysis. 
%
\bibliography{ref.bib}
\bibliographystyle{IEEEtran}

\balance

\begin{IEEEbiographynophoto} {Dr. Sunder Ali Khowaja} is currently an Assistant Professor at Department of Telecommunication Engineering, University of Sindh, Pakistan. He had the experience of working with multinational companies as Network and RF Engineer from 2008 to 2011. His research interests include Data Analytics, Computer Vision, and Artificial Intelligence for Emerging Communication Technologies.
\end{IEEEbiographynophoto}
\vskip -2\baselineskip plus -1fil
\vspace{-0.25cm}
\begin{IEEEbiographynophoto}{Kapal Dev}is currently serving as Assistant Lecturer at Munster Technological University (MTU), Ireland. He is founding chair of IEEE ComSoc special interested group titled as Industrial Communication Networks. He is serving as Associate Editor in NATURE, Scientific Reports, Springer WINE, IET Quantum Communication, IET Networks, Area Editor in Elsevier PHYCOM. 
\end{IEEEbiographynophoto}
\vskip -2\baselineskip plus -1fil
\vspace{-0.25cm}
\begin{IEEEbiographynophoto}{Parus Khuwaja } is pursuing her Ph.D. degree in financial analytics from University of Sindh, Jamshoro. She is currently working as an Assistant Professor at Institute of Business Administration, University of Sindh, Jamshoro. Her interests include Data analytics, Machine learning for Ambient Intelligence, Stock Portfolios, and Financial securities.
\end{IEEEbiographynophoto}
\vskip -2\baselineskip plus -1fil
\vspace{-0.25cm}
\begin{IEEEbiographynophoto}{Quoc-Viet Pham} [M'18] (vietpq@pusan.ac.kr) is currently working as a research professor at Pusan National University, Korea. He has been granted the Korea NRF Funding for outstanding young researchers for the term 2019–2023. He received the best Ph.D. thesis award in Engineering from Inje University in 2017. His research interests include network optimization, wireless AI.
\end{IEEEbiographynophoto}
\vskip -2\baselineskip plus -1fil
\vspace{-0.25cm}
\begin{IEEEbiographynophoto}{Nawab Muhammad Faseeh Qureshi} is an Assistant Professor at Sungkyunkwan University, Seoul, South Korea. He was awarded the 1st Superior Research Award from the College of Information and Communication Engineering based on his research contributions and performance during his studies. His research interests include big data analytics, machine learning, deep learning, and cloud computing.
\end{IEEEbiographynophoto}
\vskip -2\baselineskip plus -1fil
\vspace{-0.25cm}
\begin{IEEEbiographynophoto}{Paolo Bellavista} received MSc and PhD degrees in computer science engineering from the University of Bologna, Italy, where he is now a full professor of distributed and mobile systems. His research activities span from pervasive wireless computing to online big data processing under quality constraints, from edge cloud computing to middleware for Industry 4.0 applications.
\end{IEEEbiographynophoto}
\vskip -2\baselineskip plus -1fil
\vspace{-0.25cm}
\begin{IEEEbiographynophoto}{Maurizio Magarini} is full professor at Politecnico di Milano, Italy, His research interests are in the broad area of information and communication theory. Topics include synchronization, channel estimation, and channel coding. 
\end{IEEEbiographynophoto}

\begin{figure*}[b]
\centering
  \includegraphics[width=\linewidth]{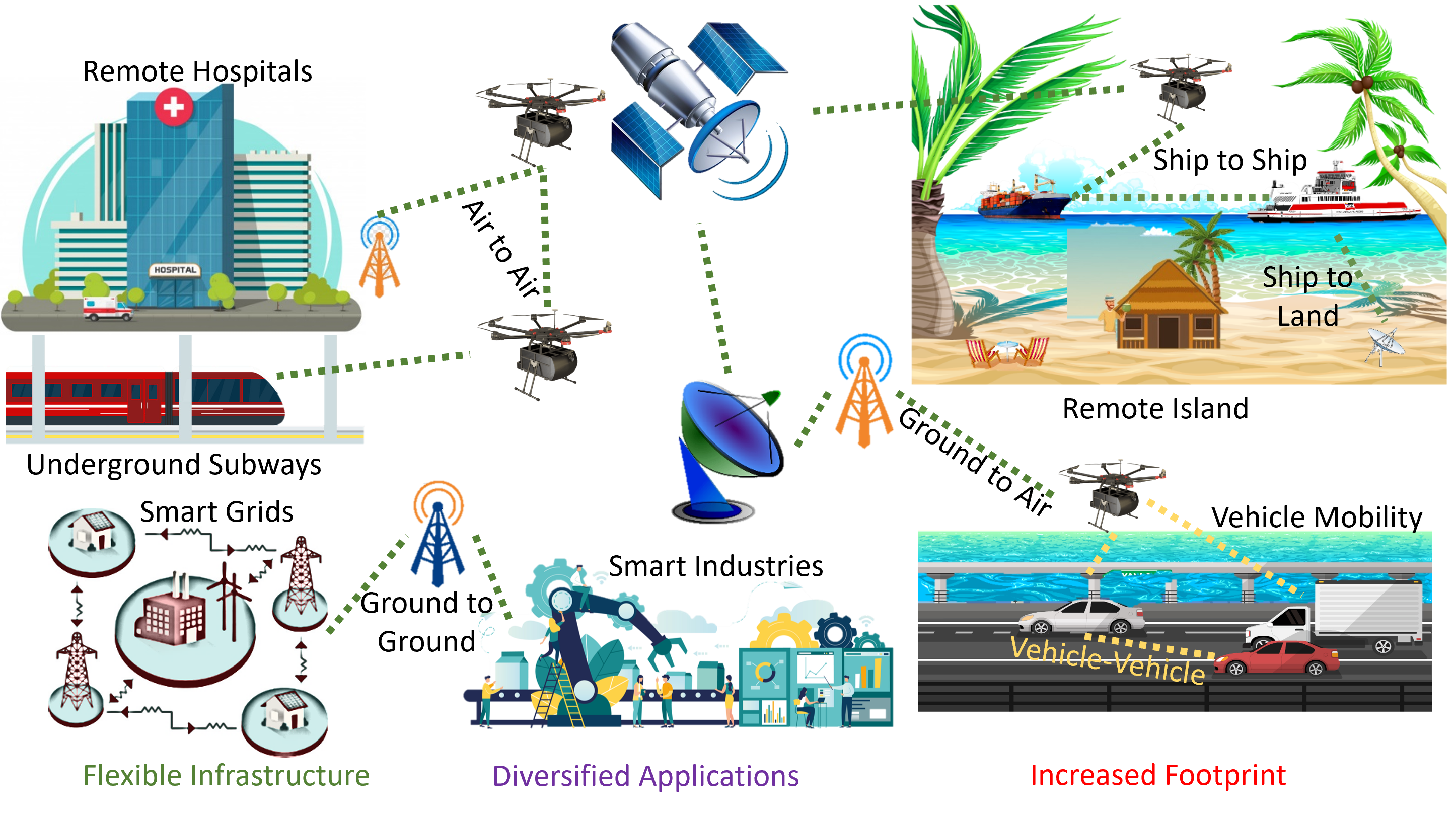}
  \caption{An overview of Next-G communication networks.}
  \label{Fig1}
\end{figure*}

\begin{figure*}[b]
\centering
  \includegraphics[width=1.07\textwidth]{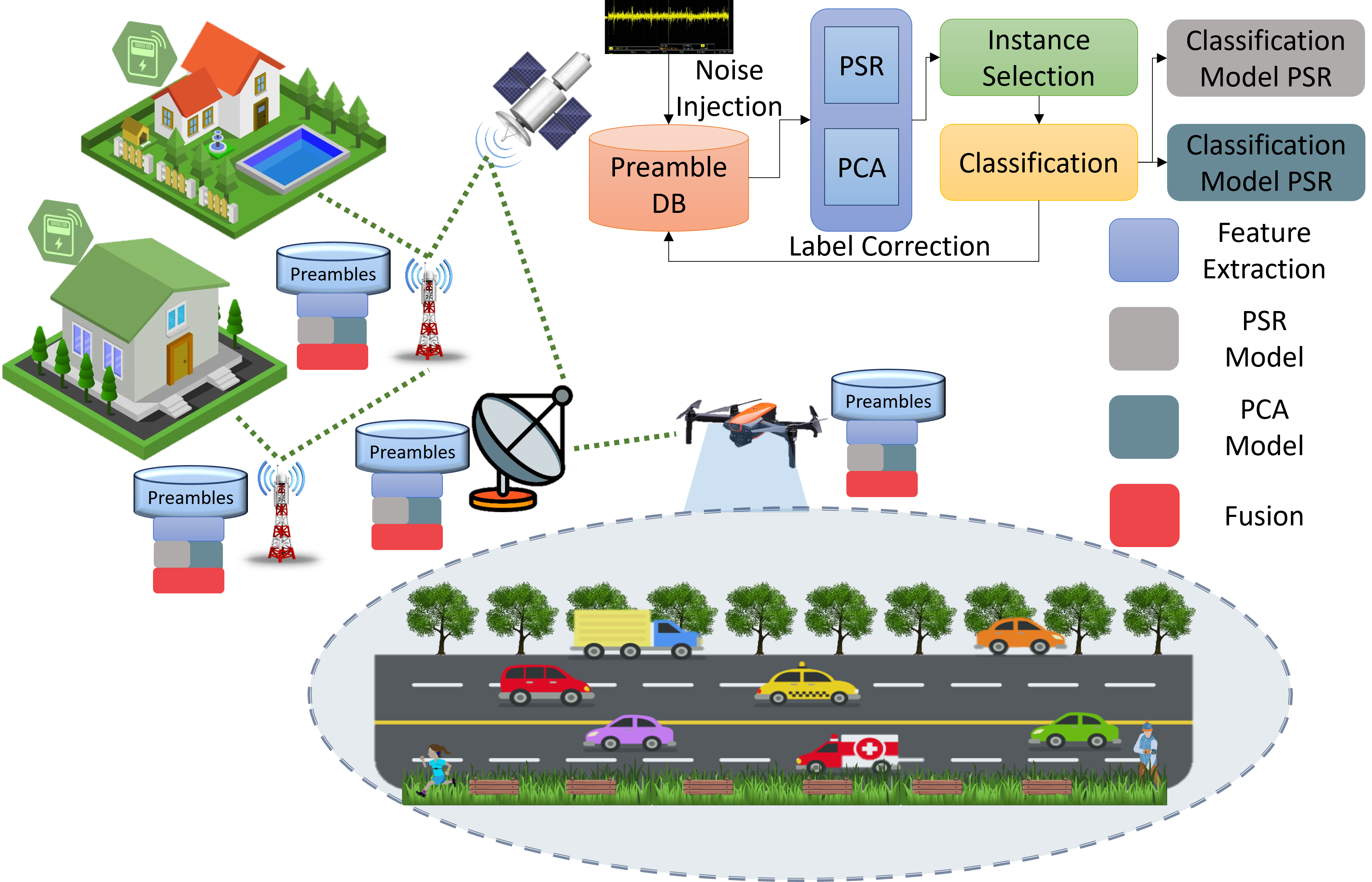}
  \caption{Proposed IIFNet framework for preamble detection in 6G and beyond communication networks.}
  \label{Fig2}
\end{figure*}

\begin{figure*}[b]
\centering
  \includegraphics[width=1.05\textwidth]{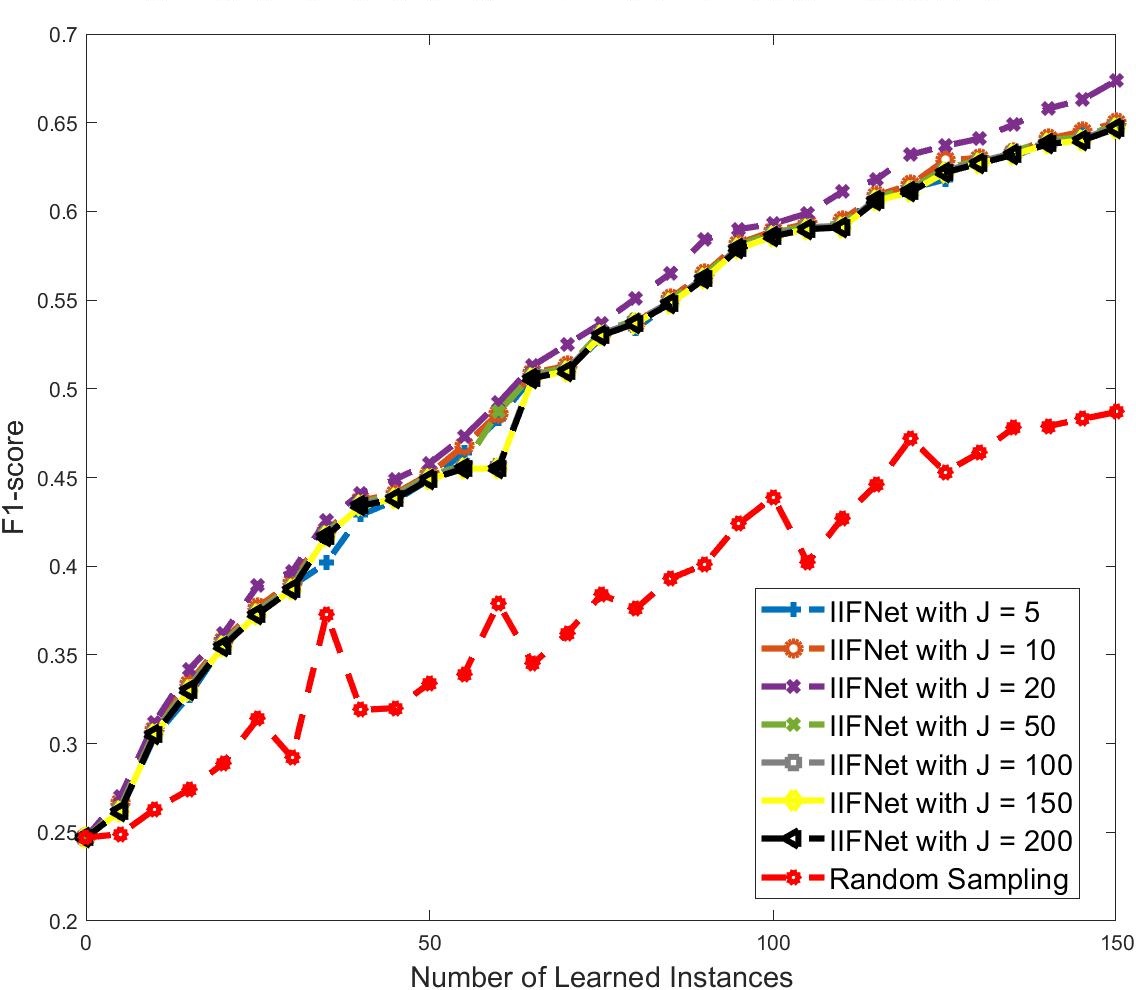}
  \caption{Results of the proposed sampling strategy in IIFNet on the collected dataset, with $J$ varying from 5 to 200. The results are obtained using PSR features trained with ELM.}
  \label{Fig5}
\end{figure*}

\end{document}